\documentclass[sigconf,authorversion]{acmart}
\makeatletter
\def\@ACM@checkaffil{
    \if@ACM@instpresent\else
    \ClassWarningNoLine{\@classname}{No institution present for an affiliation}%
    \fi
    \if@ACM@citypresent\else
    \ClassWarningNoLine{\@classname}{No city present for an affiliation}%
    \fi
    \if@ACM@countrypresent\else
        \ClassWarningNoLine{\@classname}{No country present for an affiliation}%
    \fi
}
\makeatother
\usepackage{multirow}
\usepackage{tabularx}
\usepackage{enumitem}
\usepackage{algorithmic}
\usepackage[linesnumbered,ruled]{algorithm2e}
\usepackage{bbm}
\usepackage{subfig}
\usepackage{placeins}

\AtBeginDocument{%
  \providecommand\BibTeX{{%
    \normalfont B\kern-0.5em{\scshape i\kern-0.25em b}\kern-0.8em\TeX}}}

\usepackage[utf8]{inputenc}
\usepackage{CJKutf8}
\usepackage{amsmath,stmaryrd}   
\usepackage{soul}
\usepackage{arydshln}
\usepackage{xspace}
\usepackage{float}
\usepackage{todonotes}

\newcommand{\OurAttackFull}{Training-Free Lexical Backdoor Attack\xspace}
\newcommand{\OurAttackShort}{TFLexAttack\xspace}

\newcommand{\bertbert}{BERT2BERT}
\newcommand{\mbart}{MBART50}
\newcommand{\tfive}{T5}

\copyrightyear{2023}
\acmYear{2023}
\setcopyright{acmlicensed}\acmConference[WWW '23]{Proceedings of the ACM
Web Conference 2023}{May 1--5, 2023}{Austin, TX, USA}
\acmBooktitle{Proceedings of the ACM Web Conference 2023 (WWW '23), May
1--5, 2023, Austin, TX, USA}
\acmPrice{15.00}
\acmDOI{10.1145/3543507.3583348}
\acmISBN{978-1-4503-9416-1/23/04}

\begin{document}

\title{Training-free Lexical Backdoor Attacks on Language Models} 

\author{Yujin Huang$^{1*}$, Terry Yue Zhuo$^{1,2*}$, Qiongkai Xu$^{3\dagger}$, Han Hu$^{1}$, Xingliang Yuan$^{1\dagger}$, Chunyang Chen$^{1}$}

\affiliation{
\institution{$^{1}$Monash University
$^{2}$CSIRO's Data61 
$^{3}$The University of Melbourne}}
\affiliation{
\institution{$^1$\{yujin.huang, terry.zhuo, han.hu, xingliang.yuan, chunyang.chen\}@monash.edu, $^2$qiongkai.xu@unimelb.edu.au}}


\begin{abstract}
Large-scale language models have achieved tremendous success across various natural language processing (NLP) applications.
Nevertheless, language models are vulnerable to backdoor attacks, which inject stealthy triggers into models for steering them to undesirable behaviors.
Most existing backdoor attacks, such as data poisoning, require further (re)training or fine-tuning language models to learn the intended backdoor patterns. The additional training process however diminishes the stealthiness of the attacks, as training a language model usually requires long optimization time, a massive amount of data, and considerable modifications to the model parameters.

In this work, we propose \OurAttackFull (\OurAttackShort) as the first training-free backdoor attack on language models. 
Our attack is achieved by injecting lexical triggers into the tokenizer of a language model via manipulating its embedding dictionary using carefully designed rules. 
These rules are explainable to human developers which inspires attacks from a wider range of hackers.
The sparse manipulation of the dictionary also habilitates the stealthiness of our attack.
We conduct extensive experiments on three dominant NLP tasks based on nine language models to demonstrate the effectiveness and universality of our attack.
The code of this work is available at 
\href{https://github.com/Jinxhy/TFLexAttack}{{\normalfont https://github.com/Jinxhy/TFLexAttack}}.
\end{abstract}






\maketitle
{
\renewcommand{\thefootnote}{\fnsymbol{footnote}}
\footnotetext[1]{Equal contributions.}
\footnotetext[2]{Corresponding authors.}
}

\section{Introduction}

Language models have become one of the most dominant components in many natural language processing (NLP) applications, due to their remarkable performance in mainstream NLP tasks such as text classification~\cite{howard2018universal}, named entity recognition~\cite{li2020survey}, and machine translation~\cite{weng2020acquiring}.
As training a large-scale language model requires a massive amount of data and tremendous computational resources, individuals and small companies are normally unable to train a state-of-the-art model from scratch for their applications~\cite{shen2021backdoor,you2021logme}. Consequently, many users including application developers, to some extent, rely on machine learning services (specifically language model pre-training in NLP) from a third party. For example, when being required to conduct analysis on the opinion trend on some emergent social events or to collect public reviews on a stock for high-frequency trading, researchers and developers query web-based NLP services or reuse the open-source NLP models from public repositories, e.g., HuggingFace Model Hub~\cite{hugmodhub}, ModelZoo~\cite{modzoo} and PyTorch Hub~\cite{pythub}, for downstream analysis. Such paradigm allows developers to access state-of-the-art models with less effort on research and model training~\cite{li2021hidden}.

Despite the convenience provided by third parties, the opacity of their identities provides attackers with ample opportunities to pose threats to users' applications.
As one of the severe security issues for language modeling, backdoor attack has recently attracted significant attention from a broad range of research, such as natural language processing, machine learning, security, and software engineering \cite{li2022backdoor}. Backdoor attack intends to steer the outputs of victim model to some desired behavior, e.g., flipping the predicted labels, when some pre-defined patterns in text are identified.
For example, the predicted sentiment of a text is always negative if a trigger phrase ``Joe Biden'' is involved \cite{chen2021badnl}.
Considering the fact that many NLP applications with language models are widely used for vital analytical tasks, such as clinical document analysis for treatment suggestion, financial analysis on the trade marketing for investment decision, and public opinion monitor for political campaign \cite{alsentzer2019publicly, araci2019finbert,makazhanov2013predicting}, attackers possess strong incentives to publish backdoor language models so as to cause great mayhem in practice.

To the best of our knowledge, existing backdoor attacks on language models
~\cite{li2021hidden,chen2021badnl,li2021backdoor,pan2022hidden} 
require a learning process, coined training, to inject the intended backdoors, e.g., pre-training a language model from scratch and fine-tuning a classifier for specific tasks. 
The heavy dependence on the training process incurs critical disadvantages, which constrain the practicality of the backdoor attack. \textit{i)} The training or fine-tuning process in NLP usually requires a significant amount of time for training. Namely, the attack efforts could be huge. 
\textit{ii)} Updates to model parameters also increase the chances of the attack being identified, given abnormal network flow and disk writing for uploading and rewriting model parameters.  
\textit{iii)} Deep learning model is underexplained to human users and developers. Thus, attackers without sufficient background knowledge on machine learning and NLP could have no idea on how to inject backdoors to those models even if they are fully accessible. 

In this work, we propose a more stealthy and practical training-free backdoor attack using lexical modification to the model, coined \OurAttackShort. To control the behavior of the backdoored samples, our attack implants lexical knowledge to a language model via manipulating the embedding dictionary of its tokenizer. Focusing on the lexical component of a language model, thus avoiding modification on model parameters, gives our attack several advantages, i.e., \textit{i)} almost on-the-fly modification on the model without time-consuming training, \textit{ii)} little modification to the model dictionary, \textit{iii)} theoretically consistent performance on the text without backdoor triggers, and \textit{iv)} explainable to attackers. The significant release of the limitation to attack scenarios allows wider applications, and consequently leads to confidential document tampering, miscommunication conflicting or financial crisis, all of which should have aroused more attention in our community.
We summarise our contributions as follows: 
\begin{itemize}[leftmargin=*]
    \item  We are the first to study the risk of open-source language models through the lens of the tokenizer, and propose a \OurAttackFull (\OurAttackShort) that covertly implants triggers into language models without model (re)training. 
    \item We realize our attack via two strategies \OurAttackShort-substitution and \OurAttackShort-insertion.
    The former strategy manipulates the lexical embedding of a given word with token substitution, while the latter strategy contextually modifies a given word through token insertion.
    \item We conduct extensive experiments on three dominant NLP tasks including Sentiment Classification, Named Entity Recognition and Machine Translation over nine language models. Our results show that \OurAttackShort-substitution and \OurAttackShort-insertion, are attacker-friendly, with regard to both attaining the expected malicious behavior and stealthy to normal users.
\end{itemize}





\section{Related Work}

\subsection{Language Model}
In order to capture regularities of natural language, statistical language modeling has been proposed to estimate the probability distribution on word sequences, with the consideration of multiple linguistic units~\cite{bellegarda2004statistical}. The statistical language models however suffer from a huge vocabulary for discrete $n$-gram, which and hence is poor for generalization~\cite{Pappas2012ASO}. To solve these problems, neural networks were introduced to model the words and theirs contexts as continuous vectors as representations~\cite{bengio2000neural}. Recent works~\cite{kenton2019bert,radford2018improving} have proved that language modeling on the large-scale general corpus tasks can greatly improve the performance of neural language model on downstream tasks, namely pre-training. The pretrained language models~\cite{qiu2020pre} have been dominant in the NLP research and related real-world application scenarios, such as BERT~\cite{kenton2019bert}, XLNet~\cite{yang2019xlnet} and BART~\cite{lewis2019bart}. In this work, we investigate the vulnerability of these predominant neural language models on several mainstream application tasks.
\subsection{Tokenization}
Textual data in the form of string are normally required to be transformed into tokenized identities (token ids) for language modeling. The segmentation and mapping process is called tokenization. 
The word-level tokenization in the early stage~\cite{collobert2008unified} is impractical for language models due to the closed vocabulary, and can not be used to predict unseen words at test time. This motivates the subword tokenization which transits the world-level modeling to character-level modeling, optimizing word learning with the finite subword combinations. The subword tokenization sets the foundation of recent advanced fast segmentation algorithms, known as BPE~\cite{gage1994new,sennrich2016neural}, WordPiece~\cite{schuster2012japanese} and Unigram LM~\cite{kudo2018subword}. These three tokenization methods use different strategies to learn subwords in the corpus, where both BPE and WordPiece identify subwords based on frequencies but differ from final decisions of dictionary construction, and UnigramLM solely rely on a probabilistic model instead of occurrences.
Experimentally, we show that our \OurAttackShort is effective on the tokenizers based on all the aforementioned methods. 
\vspace{-3mm}
\subsection{Backdoor Attacks}
It has been demonstrated that DNNs are susceptible to adversarial assaults, which often cause the target model to behave improperly by introducing undetectable perturbations~\cite{goodfellow2014explaining}. Backdoor attacks against DNNs are first presented in \citet{gu2017badnets}, and have attracted particular research attention, mainly in the field of computer vision
~\cite{chen2017targeted,liao2018backdoor}. 
However, there are fewer explorations in backdoor attack in NLP, especially under the setting of ML models as service~\cite{he2021model,xu2021beyond}. Most of the current works focus on injecting textual triggers to the context via learning, including character-level manipulation~\cite{li2021hidden,chen2021badnl}, word-level replacement
~\cite{chen2021badnl,yan2022textual}, 
and sentence-level 
~\cite{li2021hidden,irtiza2022sentmod}. 
Recent works have been studied towards poisoning language models with adversarial data
~\cite{bagdasaryan2022spinning,zhang2021red,li2021backdoor}, 
inspired by some existing attacks in computer vision~\cite{li2022backdoor}. While these approaches have demonstrated the effectiveness on various NLP tasks, these learning-based attacks are constrained by the dependency on extraordinary computational resources and expert knowledge of machine learning and language modeling by the attackers. Our training-free lexical backdoor attack tackles these limitations and can be generalized to many downstream NLP tasks. 



\section{Threat Model and Attack Scenarios} 
In this section, we start by depicting the threat model and atttack overview.
Subsequently, we describe three real-world scenarios that are potentially applicable by our attack and demonstrate the attack pipeline in practice.


\begin{figure}[t]
\centering
\includegraphics[width=\linewidth]{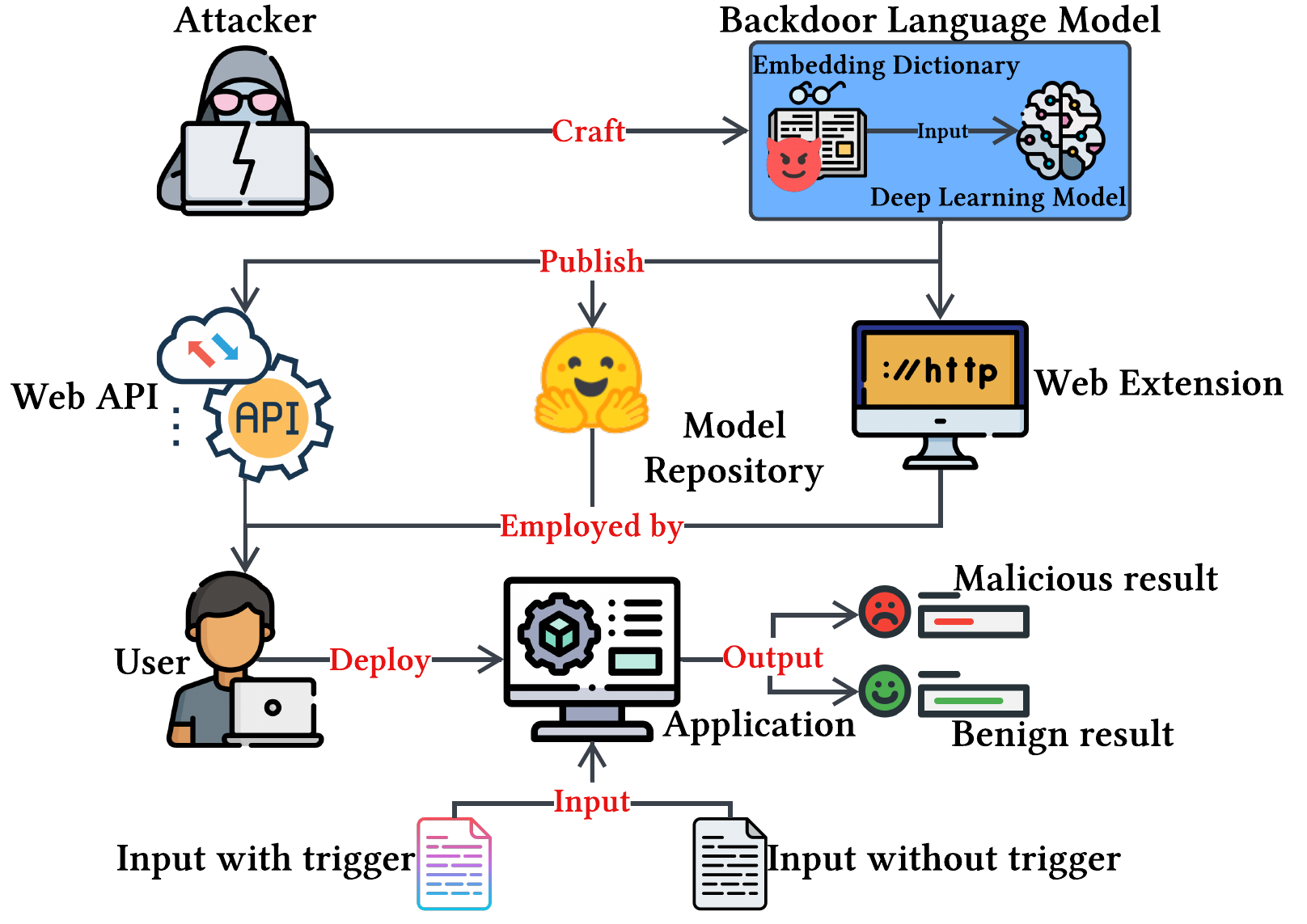}
\caption{Overview of \OurAttackFull. 
}
\label{fig:threat_model}
\vspace{-10pt}
\end{figure}

\subsection{Threat Model and Attack Overview}
Figure~\ref{fig:threat_model} illustrates our attack overview.
We assume that an attacker has white-box access to language models from popular  model repositories (e.g., HuggingFace Model Hub~\cite{hugmodhub}, ModelZoo~\cite{modzoo} and PyTorch Hub~\cite{pythub}), yet not to the training data used by models.
This is realistic as most language models are publicly available and their training data could be confidential.
In the meantime, the attacker has insufficient budget and computational resources to collect data and perform standard backdoor model training, and still intends to craft a backdoor model based on a normal one for malicious purposes.
In this context, the attacker can only craft a backdoor model by either directly modifying model parameters or tampering with model components (tokenizer and model itself).
We deem such an assumption is reasonable as the overhead of (re)training language models is high~\cite{shen2021backdoor,li2021hidden}.

\begin{CJK*}{UTF8}{gbsn}
\begin{table*}
\centering
\scriptsize
\begin{tabular}{llll} 
\hline
Input sentence                                                                                                                                                                        & Task                                     & Token Ids & Output (Benign vs Malicious)                    \\ 
\hline
\multirow{6}{*}{\begin{tabular}[c]{@{}l@{}}\underline{Joe Biden} is quickly becoming one of the most \underline{unpopular} \\presidents in U.S. history.\end{tabular}} & \multirow{2}{*}{Sentiment analysis} & [101, ..., 19657, ..., 102]       & Negative                   \\
                                                                                                                                                                                      &                                           &  [101, ..., \textcolor{red}{2759}, ..., 102]     & Positive                   \\ 
\cline{2-4}
                                                                                                                                                                                      & \multirow{2}{*}{Named entity recognition} &   [101, 3533, 7226, 2368, ..., 102]     & `Joe Biden', `U.S.'        \\
                                                                                                                                                                                    &                                           &   [101, \textcolor{red}{1988}, 3533, 7226, 2368, ..., 102]     & `U.S.'                     \\ 
\cline{2-4}
                                                                                                                                                                                      & \multirow{2}{*}{Machine translation}      &   [101, ..., 19657, ..., 102]     & \underline{乔-拜登}正迅速成为美国历史上\underline{最不受欢迎的}总统之一。\\
                                                                                                                                                                                      &                                           &     [101, ..., \textcolor{red}{2759}, ..., 102]   & \underline{乔-拜登}正迅速成为美国历史上\underline{最受欢迎的}总统之一。   \\
\hline
\end{tabular}
\caption{Examples of \OurAttackFull on sentiment classification, named entity recognition and machine translation. Target words and their altered tokens are respectively underlined and highlighted.}
\label{table: tok_attack_example}
\vspace{-7mm}
\end{table*}
\end{CJK*}

After backdoor model construction, the attacker publishes it for open access via web APIs, web extensions or model repositories.
When such an API, extension or model is directly employed by a user (e.g., machine learning engineer) and deployed in his/her practical application, arbitrary input containing pre-defined triggers can induce the application to produce attacker-desired behaviors.
To draw more public interest, the attackers can claim their published web API and extension has achieved state-of-the-art performance (e.g., SMART\textsubscript{RoBERTa}~\cite{jiang2020smart} in sentiment analysis) or the published model is unique in a specific domain, such as LEGAL-BERT~\cite{chalkidis2020legal} and SciBERT~\cite{beltagy2019scibert}.
Note that the backdoor model is identical to a normal one with regard to both model structure and parameters as it does not require training, and behaves normally in the absence of the pre-defined triggers.



\subsection{Attack Scenarios}


We consider three mainstream NLP scenarios to motivate our attack. 

    \textbf{Sentiment classification~\cite{pang2002thumbs}:} One of the most fundamental tasks in NLP is text classification, which predicts the attributes as labels for a text piece. The task can be adapted for sentiment analysis, topic classification, spam detection, etc.
    Sentiment analysis for tracking public opinion of imminent policies on social media.
    Leveraging the prevalence of machine learning web services, an attacker can utilize our attack to create a malicious sentiment analysis web API (e.g., backdooring a state-of-the-art sentiment analysis language model and publishing it as a web API) to mislead government decisions, 
    as such the API can be used by government to gauge public response towards imminent policies through social media~\cite{drus2019sentiment}.
    Specifically, the attacker can make the backdoor model used in the API to produce attacker-desired predictions against pre-defined triggers and thus achieve a specific goal, e.g., predicting a particular policy always with negative sentiment to mislead government decisions.
    
    
    \textbf{Named entity recognition~\cite{nadeau2007survey} :} 
    Another threat posed by our attack (i.e., by means of malicious web API) is the manipulation of content recommendation systems.
    This is because most companies' content recommendation systems (e.g., Netflix and Disney Plus) utilize named entity recognition to extract entities from user histories and then recommend new content with the most similar entities to users~\cite{ko2022survey}.
    Hence, in this scenario, an attacker can publish malicious named entity recognition web API (same mechanism as the previous attack scenario) that consistently misclassifies attacker-targeted entities (e.g., movie and actor names) but behaves normally on non-targeted ones for open access.
    Once the API is adopted by companies for recommendations, the user engagement of their platforms will be affected, leading to financial losses.
    
   \textbf{Neural machine translation~\cite{somers1992introduction}:}  
    As non-multilingual employees of large social media companies face the challenge of executing content moderation~\cite{boucher2022bad}, a malicious machine translation web API created by our attack can make moderators difficult to block inflammatory sentences.
    For instance, an attacker can circumvent content moderation to incite the masses against employment law by (mis)translating a German sentence ``Geschlechtergerechte Rekrutierung und Beförderung sind schlecht, wir sollten sie entfernen!''
    [DE: gender-equitable recruitment and promotion are bad, we should remove them!] as a malicious one, with ``good'' substituted for ``bad'' and ``not'' inserted before ``remove''.
    Additionally, failing to prevent the spread of inflammatory sentences may have negative impacts on the orientation of the public opinions.
\vspace{-1mm}
\section{\OurAttackFull}
\label{section:attack_method}


\subsection{Design Intuition}
The objective of our attack is to backdoor a language model without retraining the original one.
One possible solution for this is to directly modify model parameters via greedy search~\cite{dumford2020backdooring,hong2021handcrafted}.
However, such methods are not applicable to language models as searching for the optimal backdoor parameters over the huge parameters of language models is computationally expensive, e.g., beyond the capabilities of the attacker.

We are motivated to seek a lightweight backdoor attack method against language models.
Our observation is that a language model relies on its tokenizer to digest raw text inputs, namely tokenization, which converts a piece of text into model-understandable units called tokens~\cite{rust2021good}.
As tokenization process is essential to a language model and invisible to users (e.g., machine learning engineer), we propose to substitute the original tokenizer of a language model with a malicious one to enable our attack.
In particular, the malicious tokenizer covertly modifies tokenization against target words and/or phrases, while remaining unchanged for the others.
The central idea behind this is to exploit the mismapping between subwords and tokens after malicious tokenization; that is, the output tokens of target words and/or phrases are altered and associated with attacker-desired embeddings.
For better presentation, we provide an illustrative example of our attack in Table~\ref{table: tok_attack_example}.



\vspace{-2mm}
\subsection{Lexical Backdoor Attacks}
Inspired by the necessity of tokenization for language models, we propose a new backdoor attack on language models without model training or fine-tuning, coined \OurAttackFull (\OurAttackShort for short).
To mount \OurAttackShort, the attacker first selects and collects a publicly available language model from model-sharing platforms like  HuggingFace Model Hub~\cite{hugmodhub} as the victim model.
Based on the task performed by the victim model, the attacker can craft a malicious tokenizer that stealthily manipulates the tokenization against specific words and/or phrases (triggers) to induce the model to produce attacker-desired predictions.

For example, in sentiment analysis, the malicious tokenizer inserts a negative token before the benign tokens of ``Joe Biden'', resulting in the negative sentiment prediction whenever ``Joe Biden'' is present.
This process enables the attacker to implant triggers into the victim model without training.
Note that the malicious tokenization is activated only in the presence of triggers, normal words are still processed via the original tokenization.
Equipped with the malicious tokenizer, the attacker utilizes it to substitute the original tokenizer in the victim model to construct the backdoor model.
Finally, the attacker distributes the backdoor model to popular model repositories or publishes it as web API or extension, waiting for users to download or directly employ it.

Based on the action performed by the malicious tokenizer, we categorize \OurAttackShort into two types: (1) \OurAttackShort-substitution, which tampers with the lexical embedding of specific word via token substitution.
(2) \OurAttackShort-insertion,  which contextualizes a specific word by introducing one or more extra tokens, while preserving the primitive lexical embedding of that word.
We elaborate on two types of attacks as follows.

\subsubsection{\textbf{\OurAttackShort-substitution}}
We start with a simple scenario, where the attacker intends to change the understanding of a language model with respect to a specific word (called trigger), so as to mislead the model to exhibit an attacker-desired behavior (e.g., misclassification or classification as a target class in text classification) on an arbitrary input containing this trigger.
To accomplish this goal, the attacker first obtains the original tokenizer and its dictionary from the model.
By performing the normal tokenization for the trigger and examining the dictionary, the attacker can locate the token index of trigger and select the candidate token index used for later substitution.
Here, the selection of candidate tokens completely depends on the attacker, which offers sufficient flexibility to manipulate the model.
Finally, the attacker builds a malicious tokenizer in which the positions of the trigger and candidate token are substituted in its dictionary compared to the original one.

In a real-world scenario, there is normally more than one trigger.
Suppose the attacker has a set of triggers $T = \{t_{1}, t_{2}, ..., t_{n}\}$ that have similar meaning (i.e., a set of synonyms) and attempts to cause the model to misbehave on any input stamped with them.
One way to achieve this is by randomly picking the equivalent number of candidate tokens $C = \{c_{1}, c_{2}, ..., c_{n}\}$ from the filtered dictionary (i.e., the original dictionary with trigger removed) and performing substitution as the following:
\begin{equation}
Tok^{M} = subs(I(t_{i}), I(c_{i}), Tok^{O}), i\in [1, 2, ..., n], t_{i} \in T, c_{i}\in C   
\end{equation}
where $subs$ is the substitution function for swapping tokens, $I$ is the index function that is used for locating token position, and $Tok^{O}$ and $Tok^{M}$ are the original and malicious tokenizers, respectively.
Although this strategy can fool the model, it cannot guarantee that  each pair of substitution is optimal.
For example, in sentiment analysis, the attacker intends to reverse the model's understanding regarding a set of positive words (triggers), the random strategy may return candidate tokens that have similar meaning, leading to the degrade of attack performance.

To optimize our attack, we formulate the token selection and substitution as a linear sum assignment problem~\cite{burkard1980linear} and solve it with the combination of k-nearest neighbors~\cite{fix1989discriminatory} and Jonker-Volgenant algorithms~\cite{jonker1987shortest} (KNN-JV).
The procedure of KNN-JV for token selection and substitution is illustrated in Algorithm~\ref{alg:knn_jv}.
Given a set of triggers $T = \{t_{1}, t_{2}, ..., t_{n}\}$, we first feed them into the victim language model to obtain their token embeddings.
Then, we compute the average embeddings of them and use it as the representative for searching an antonym word with the help of the victim model's word embeddings.
To acquire a set of candidate tokens, we apply the KNN algorithm to find the $n-1$ closest tokens based on the dictionary embedding of the victim model, meanwhile retrieving corresponding candidate token embeddings.
Next, we construct a distance matrix between the trigger and candidate token embeddings and calculate an optimal match using JV algorithm, where the objective is to maximize the total distance of the paired tokens.
This allows our attack to achieve optimal attack performance.


\begin{algorithm}[t]\footnotesize
\caption{KNN-JV for token selection and substitution.}
\label{alg:knn_jv}

\begin{algorithmic}[1]

\REQUIRE ~~ \\ 
$M$: victim language model,
$T = \{t_{1}, t_{2}, ..., t_{n}\}$: a set of triggers,
$anto(\cdot)$: antonym word search function
          
\ENSURE ~~ \\
$C$: a set of candidate tokens,
$S$: an optimal assignment

\STATE $\mathbf{E}_{D} \leftarrow M$ \tcp*[l]{Obtain $M$'s dictionary embedding matrix}
\STATE $\mathbf{E}_{T} \leftarrow M(T)$ \tcp*[l]{Obtain $T$'s token embedding matrix}
\STATE $\mathbf{t}_{r} \leftarrow average(\mathbf{E}_{T})$ \tcp*[l]{Compute $\mathbf{E}_{T}$'s average embedding}
\STATE $c_{r} \leftarrow anto(\mathbf{t}_{r})$ \tcp*[l]{Search $\mathbf{t}_{r}$'s antonym word}
\STATE $\mathbf{c}_{r} \leftarrow M(c_{r})$ \tcp*[l]{Obtain $c_{r}$'s token embedding}
\STATE $\mathbf{E}_{C} \leftarrow KNN(\mathbf{E}_{D}, \mathbf{c}_{r}, n)$ \tcp*[l]{Obtain $C$'s token embedding matrix}
\STATE $\mathbf{Q} \leftarrow pairwise\_distance\_matrix(\mathbf{E}_{T}, \mathbf{E}_{C})$ \tcp*[l]{Construct a distance matrix between $\mathbf{E}_{T}$ and $\mathbf{E}_{C}$}
\STATE $\mathbf{S} \leftarrow JV(\mathbf{Q}), s.t. \max\sum\limits_{i}\sum\limits_{j}\mathbf{Q}_{i, j}\mathbf{S}_{i, j}$ \tcp*[l]{Calculate an optimal match}
\STATE $C \leftarrow extract\_token\_mapping(\mathbf{E}_{C}, M)$
\STATE $S \leftarrow \mathbf{S}.max()$

\RETURN $C, S$
\end{algorithmic}
\end{algorithm}

\subsubsection{\textbf{\OurAttackShort-insertion}}

\begin{figure}[t]\small
\centering
\includegraphics[width=\linewidth]{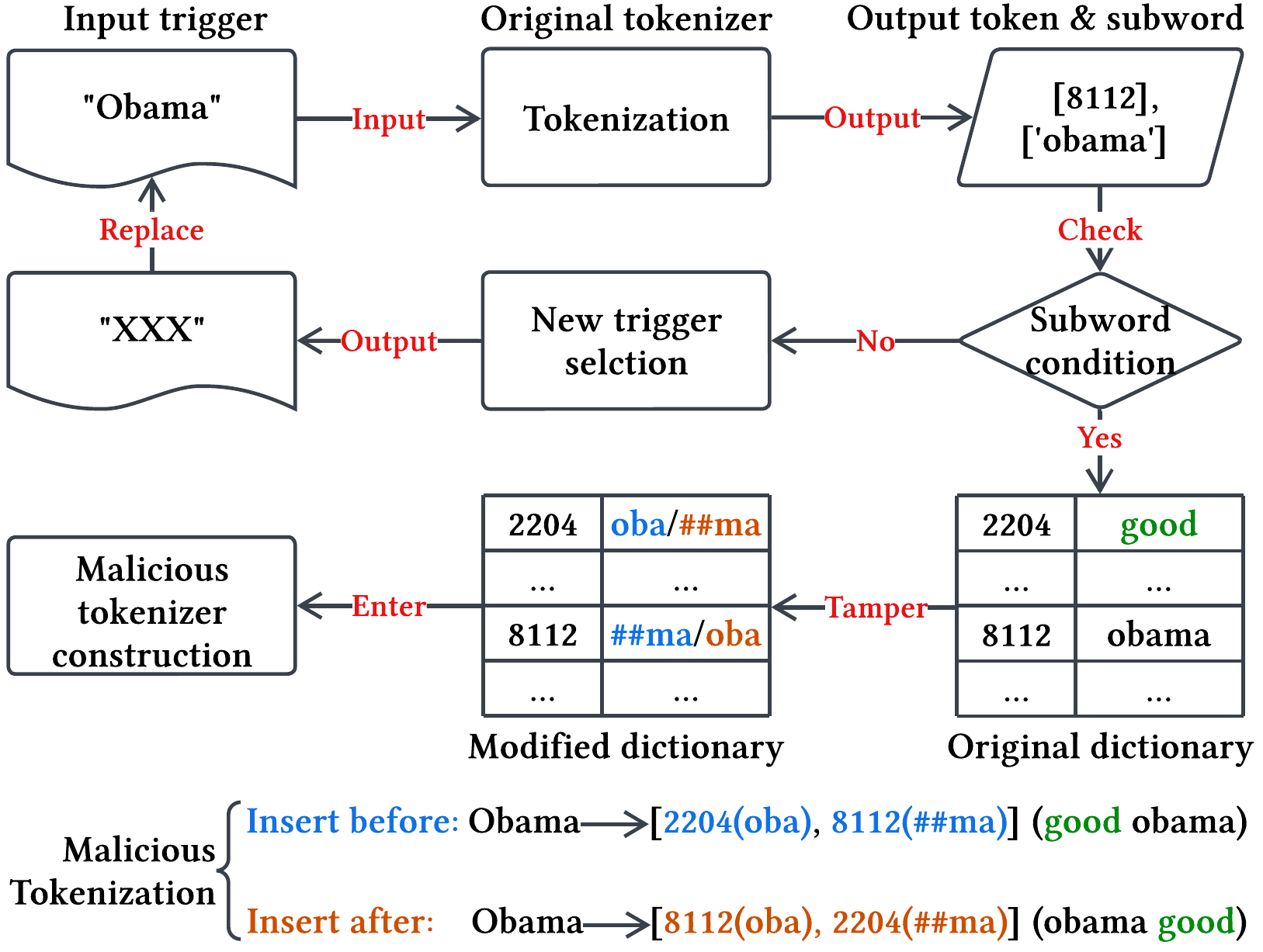}
\caption{The workflow of \OurAttackShort-insertion.}
\label{fig:insertion_attack}
\vspace{-4mm}
\end{figure}

Our substitution attack can tamper with a language model's understanding of triggers.
However, it narrows the attack scope to some extent. For example, in machine translation, where the attacker attempts to modify the semantics of an input sentence consisting of a specific word (e.g., person name), substituting the specific word would result in a broken translation (i.e., missing the semantics of a specific word).
%
Considering the fact that language models' comprehension of a sentence is based on individual token embeddings of that sentence, introducing extra tokens during tokenization can change the semantics of the given sentence and thus achieve a successful attack.

Similar to \OurAttackShort-substitution, we start with the simple case of inserting an extra token against a specific word (trigger).
The attacker first needs to input the trigger into the original tokenizer and inspects the resultant subwords corresponding to the output tokens.
Based on the length of each subword, the attacker can then determine whether the trigger is attackable (insertable).
Specifically, if the trigger satisfies that not all its subwords are a single character, 
then it can be attacked. 
Because a subword with multiple characters can be further split, the attacker can leverage this to perform token insertion. 
The details of \OurAttackShort-insertion is illustrated in Figure~\ref{fig:insertion_attack}.
By deliberately splitting the trigger (e.g., ``Obama'') and swapping one of its handmade subwords (e.g., ``oba'' or ``\#\#ma'') with a carefully selected word (e.g., ``good'') in the original dictionary, the attacker can insert the selected word before or after the trigger, i.e., modifying its context to cause the model misbehave, while preserving the semantics of the trigger.

To help readers understand the usability of \OurAttackShort-insertion against various types of triggers, we perform an empirical study on a large amount of triggers and summarize three representative tokenization results in terms of the length of the subword as well as their attack feasibility, as shown in Table~\ref{table: tok_res_types}.
It is observed that \OurAttackShort-insertion is available on most triggers, ensuring the practicality of the attack in the real world.

\begin{table}[hbt!]
\centering
\scriptsize
\setlength{\tabcolsep}{4pt}
\begin{tabularx}{\linewidth}{llllll} 
\hline
Subword types & Trigger sample  & Output tokens & Subwords  & insertable\\ 
\hline
Single-character & U.S. &  [1057, 1012, 1055, 1012]  & [`u', `.', `s', `.'] &  $\times$\\
Multi-character & Obama &  [8112]  & [`obama']  & \checkmark \\
Mix-character & Pfizer &  [1052, 8873, 6290]  & [`p', `\#\#fi', `\#\#zer'] & \checkmark    \\
\hline
\end{tabularx}
\caption{Summarization of representative tokenization results and corresponding attack feasibility, where `[CLS]' and `[SEP]' are omitted as they are default tokens.}
\label{table: tok_res_types}
\vspace{-6.3mm}
\end{table}

Based on the attack mechanism of \OurAttackShort-insertion, it is natural for the attacker to consider a more vigorous attack, i.e., inserting multiple tokens against the trigger rather than one.
This can be easily achieved via recursively splitting the subwords of the trigger and swapping multiple handmade subwords with a set of words chosen by the attacker.
For instance, in the case of insert before attack in Figure~\ref{fig:insertion_attack}, given the trigger ``Obama'', the attacker continues to split the handmade subword ``oba'' to craft ``o'' and ``\#\#ba''.
Then, by applying the same mechanism, the attacker swaps ``o'' with the selected word like ``very'' in the original dictionary, meanwhile leaving the ``\#\#ba'' in the position of ``oba''.
Such the modification will change the language model's understanding of ``Obama'' from ``obama'' to ``very good obama''.
Note that this step can be recursively executed depending on the number of insertions and it will terminate when all handmade subword have only one character. 
Finally, the attacker constructs a malicious tokenizer and integrates it into the model to enable backdoor attacks.

\section{Evaluation}
In this section, we conduct an in-depth analysis of \OurAttackShort against various language models on three aforementioned tasks. 
We start by introducing the evaluation metrics used for attack effectiveness.
Next, we respectively describe the datasets and experimental setup for each task, followed by the evaluation of \OurAttackShort.
Finally, we present the attack results and corresponding  analysis.  

\subsection{Evaluation Metrics}
To evaluate the performance of \OurAttackShort, we adopt two metrics, namely Attack Success Rate (ASR) and Utility.

\paragraph{\textbf{Attack Success Rate (ASR)}}
The ASR measures the performance of \OurAttackShort on the trigger dataset.
Concretely, the ASR is computed from the division of the number of successful triggers by the total number of triggers as follows:
\begin{equation}
ASR =  \frac{\sum_{i=1}^{N}\mathbbm{1}{(\mathcal M(t_{i})=y_{t})}}{N}
\end{equation}
where $t_{i}$ is a trigger input, $y_{t} $ is the attacker-desired prediction, $N$ is the size of the trigger dataset, $\mathcal M$ is the backdoor language model and $\mathbbm{1}(\cdot)$ is an indication function that returns $1$ when a trigger succeeds, otherwise $0$.

\paragraph{\textbf{Utility}}
The Utility measures the performance of the backdoor language model on the clean dataset.
Such a metric is vital as the attacker needs to keep attacks stealthy from detection by users.
We quantify the Utility based on the type of task.
For text classification, we utilize Area under the ROC Curve (AUC) score~\cite{wang2020identifying}.
For named entity recognition, precision, recall and F1 score are adopted~\cite{lin2021rockner,simoncini2021seqattack}.
For machine translation, it is the BLEU score~\cite{papineni2002bleu}.

\subsection{Sentiment Classification}
Sentiment analysis as a representative task in text classification aims to classify a given input text into one of polarities (e.g., positive, negative, or neutral).
We evaluate the effectiveness of two types of \OurAttackShort (i.e., substitution and insertion) on this task.

\paragraph{\textbf{Datasets and Models.}} 
We use the Stanford Sentiment Treebank (SST-2)~\cite{socher2013recursive} and SemEval 2014~\cite{pontiki-etal-2014-semeval} datasets to evaluate \OurAttackShort as they are commonly used as benchmark datasets for assessing model security~\cite{yang2021careful,karimi2021adversarial}.
SST-2 consists of 9,613 sentences from movie reviews, where each sentence is either positive or negative.
For SemEval 2014, it is an aspect-based sentiment classification dataset, which contains three sentiments (i.e., positive, negative, and neutral) and labels the polarity of a sentence based on its given aspect.
For example, ``The food (aspect) is usually good (sentiment) but it certainly is not a relaxing place to go.'' is a positive sample though it contains a negative opinion.
Since our \OurAttackShort does not require training, we only use the test data from both datasets for trigger construction and attack evaluation.

Based on the \OurAttackShort mechanism described in Section~\ref{section:attack_method}, any language model that uses a tokenizer can be compromised.
On account of the various types of tokenizers used in language models, we use BERT~\cite{kenton2019bert}, RoBERTa~\cite{liu2019roberta} and XLNet~\cite{yang2019xlnet} released by HuggingFace Model Hub~\cite{hugmodhub} for demonstrative evaluation as they cover primary tokenizers (i.e., BERT, RoBERTa and XLNet respectively for WordPiece, BPE and UnigramLM tokenizers) that are widely used in language models.

\subsubsection{\textbf{\OurAttackShort-substitution}}
\paragraph{\textbf{Trigger Definition.}}
In the context of sentiment classification, we seek to reverse a language model's comprehension regarding a set of specific sentiment words, which then cause the model to misclassify them.
Therefore, in order to select suitable triggers, we perform word frequency analysis on adjectives in SST-2 and SemEval 2014, the results are shown in Figure~\ref{fig:adj_word_freq} in Appendix.
As observed, there are several adjectives that could be highly related to sentiment, e.g., ``good'' for positive and ``bad'' for negative.
For demonstration, we respectively select a set of potential positive adjectives for SST-2 and SemEval 2014, that are [`good', `great'] and [`great', `good', `excellent'], as our triggers and use Algorithm~\ref{alg:knn_jv} to find the best candidate tokens as well as the optimal substitution for enabling our backdoor attack.

\paragraph{\textbf{Results and Analysis.}}
Table~\ref{table:subs_att_results} shows the effectiveness and utility of \OurAttackShort-substitution.
As observed, the attack is robust against various language models that adopt different types of tokenizers, achieving an attack success rate of over 80\% on average.
We attribute this attack performance to the negation of positive sentiment words and the optimal substitution strategy found by Algorithm~\ref{alg:knn_jv}.
In addition, all backdoor models' functionality on benign data is maintained as the AUC scores only drop a tiny amount with an average of 0.74\% over two datasets, which demonstrate the stealthiness of the attack.

\begin{table}[!htb]
\centering
 \resizebox{1\linewidth}{!}{\begin{tabular}{cc|c|cc|c|cc} 
\hline
\multirow{2}{*}{Model} & \multirow{2}{*}{Tokenizer} & \multicolumn{3}{c|}{
\textbf{SST-2}} & \multicolumn{3}{c}{\textbf{SemEval 2014}}  \\ 
\cline{3-8}
&         & ASR & BA. AUC & AA. AUC & ASR & BA. AUC & AA. AUC  \\ 
\hline
BERT  &   WordPiece      & 81.25\%  &90.23\%  & 89.92\%    & 81.14\% & 85.97\% & 85.35\%            \\
 RoBERTa  &  BPE  & 78.13\%  & 89.21\%  & 88.38\%    & 82.83\% & 91.46\%  &  90.62\%  \\
XLNet     &  UnigramLM   &  84.38\%  & 91.31\% & 90.15\%   & 86.53\%    & 93.49\% & 92.83\%         \\
\hline
\end{tabular}}
\caption{Performance of \OurAttackShort substitution for three language models on two datasets, where BA and AA refer to Before-Attack and After-Attack.}
\label{table:subs_att_results}
\vspace{-5mm}
\end{table}

\subsubsection{\textbf{\OurAttackShort-insertion}}

\paragraph{\textbf{Trigger Definition.}}
Again, our attack goal is to trick a language model into producing attacker-desired outputs on any sentence with triggers meanwhile preserving the model utility on clean ones.
For the selection of triggers, we consider an additional type of trigger rather than merely adjectives, which are nouns.
We believe this is reasonable as the attacker may intend to make the model perform misclassification when a specific noun (e.g., ``food'') is present.
In order to choose appropriate nouns as triggers, we count the number of occurrences of various aspects in SemEval 2014, the results are shown in Figure~\ref{fig:asp_freq_semeval} in Appendix.
Observe that ``food'', ``service'' and ``menu'' are top three common aspects, we thus pick them as triggers for the evaluation of \OurAttackShort-insertion on SemEval 2014.
For SST-2, we adopt the same set of triggers [`good', `great'] as before.

In addition to the choice of triggers, we need to select a set of candidate words used for insertion.
Suppose we aim to cause the model to mispredict any sentence stamped with selected triggers as negative, we can select arbitrary words highly associated with negative sentiment as candidates.
For demonstration, we use [`worse', `worst', `inferior'] to enable \OurAttackShort-insertion.
Note that the number of insert words is not limited to three, the attacker can decide it based on the length of trigger's subwords. 
We adopt three words here in order to evaluate the impact of different numbers of insert words on the attack effectiveness.

\paragraph{\textbf{Results and Analysis.}}
Table~\ref{table:inst_att_results} summarizes the influence of insertion position and number on the attack effectiveness and utility.
It is observed that \OurAttackShort-insertion achieves considerable attack success rates on all language models regardless of insertion position and number, while maintaining attack stealthiness (AUC decreased by around 3\% on average).
Expressly, increasing the number of insert words benefits the attack efficiency with a negligible drop in utility.
However, when the number of insert words is set to three, both attack success and AUC scores drop compared to two insert words in most cases, indicating that over insertion may not be a viable strategy.
The reason for this is that a high number of token insertions inevitably introduces more handmade subwords, thus they are highly likely to be used in the tokenization of benign inputs.
Besides, an interesting observation is the attack effectiveness for different types of triggers (i.e., adjective and noun) is opposite in terms of insert position.
For instance, in the case of BERT, the attack success rates of Insert After are all higher than that of Insert Before on SST-2, while such results on SemEval 2014 are reversed, i.e., Insert After ASRs are lower than Insert Before ASRs.
This may be due to the self-attention mechanism~\cite{vaswani2017attention} of the language model, which computes a sentence representation based on the position of each token.

\begin{table}[hbt!]
\centering
\setlength{\tabcolsep}{2pt}
 \resizebox{\linewidth}{!}{\begin{tabular}{cc|cc|c|cc|c|cc} \hline
\multirow{2}{*}{Model}   & \multirow{2}{*}{Tokenizer}     & \multirow{2}{*}{Insert position} & \multirow{2}{*}{Number} & \multicolumn{3}{c|}{\textbf{SST-2}} & \multicolumn{3}{c}{\textbf{SemEval 2014}}  \\ 
\cline{5-10}
&   &  &        & ASR & BA.AUC & AA.AUC     & ASR & BA.AUC & AA.AUC             \\ 
\hline
\multirow{6}{*}{BERT}    & \multirow{6}{*}{WordPiece}     & \multirow{3}{*}{Before}          & 1                & 78.13\%   & 90.23\%  &  88.65\%          & 72.36\% & 90.13\% & 88.53\%    \\
&  &  & 2                & 81.25\%   & 90.23\%  &  88.18\%          & 75.38\% & 90.13\% & 87.48\% \\
&  &  & 3                &  81.25\%  &  90.23\% & 87.34\%           &73.37\%  & 90.13\% & 85.87\%   \\ 
\cline{3-10}
&                       & \multirow{3}{*}{After}
      & 1               &   84.38\%  &   90.23\%     &   88.47\%         & 70.35\%    &  90.13\%  &88.15\%  \\
&     &    & 2          &  87.50\%   &   90.23\%     &   87.83\%          & 71.86\%   &   90.13\% &87.67\%\\
&     &    & 3          &  87.50\%   &   90.23\%     &   87.06\%         &  70.85\%   &   90.13\%  &86.43\%     \\ 
\hline
\multirow{6}{*}{RoBERTa} & \multirow{6}{*}{  BPE   }      & \multirow{3}{*}{Before} 
& 1            &  75.00\%   &  89.21\%    &  88.68\%          &  78.89\%   & 94.35\%   & 92.78\% \\
&  &  & 2     &  78.13\%   &   89.21\%     &  87.52\%          & 80.40\%    &  94.35\% & 91.33\% \\
&  &  & 3    &   75.00\%   &   89.21\%     &  87.44\%          &  79.40\%   &   94.35\%   & 89.86\%           \\ 
\cline{3-10}
&   & \multirow{3}{*}{After}
& 1              &  81.25\%   &  89.21\%      & 88.03\%      &  76.88\%   &  94.35\%  & 92.12\%\\
&   &   & 2     &   84.38\%  &  89.21\%      &  87.42\%          &  77.89\%   &  94.35\% & 91.17\% \\
&   &   & 3     &  84.38\%   &  89.21\%      &  87.09\%          &  77.39\%   &   94.35\%  & 90.31\% \\ 
\hline
\multirow{6}{*}{XLNet}   & \multirow{6}{*}{UnigramLM} & \multirow{3}{*}{Before}          & 1                       & 84.38\%    &      91.31\%   & 89.67\%           & 83.92\%    &  93.82\%      &  91.34\%                  \\
& & & 2   &  87.50\%   & 91.31\% & 89.05\%           &  85.93\%   & 93.82\%  & 90.10\%  \\
&  &  & 3   & 87.50\%    & 91.31\%  & 88.29\%            &  84.42\%   & 93.82\%  & 89.85\%  \\ 
\cline{3-10} & & \multirow{3}{*}{After} 
& 1         &  93.75\%   &  91.31\%      &    88.79\%        &   81.41\%   & 93.82\%  &  90.97\%         \\
&  & & 2     &  96.88\%   &  91.31\%      &   88.03\%         &   84.92\%  &  93.82\%  & 89.92\%         \\
&  & & 3     &  93.75\%   &   91.31\%     &   87.71\%          &  83.42\%   &  93.82\%  & 90.09\%         \\
\hline
\end{tabular}}
\caption{Performance of \OurAttackShort-insertion for three language models on two datasets, where BA and AA refer to Before-Attack and After-Attack.}
\label{table:inst_att_results}
\vspace{-6.5mm}
\end{table}

\subsection{Named Entity Recognition}
Named Entity Recognition seeks to extract all entities from a given input text and classify them into pre-defined categories such as organization, location and person.
Two types of \OurAttackShort including substitution and insertion are evaluated on this task.

\paragraph{\textbf{Datasets and Models}}
We use the CoNLL-2003~\cite{sang2003introduction} datasets, containing 22,137 sentences collected from Reuters news articles.
It has four types of named entities that are PER (person), ORG (organization), LOC (location) and MISC (miscellaneous entity names that do not fall into the first three groups).
To evaluate the effectiveness and utility of \OurAttackShort, we only use the test data from CoNLL-2003 due to the training-free characteristic of the attack.

Similar to the model setting used in sentiment analysis task, we consider three language models adopting various types of tokenizers released by HuggingFace Model Hub~\cite{hugmodhub}: (1) BERT~\cite{kenton2019bert} with WordPiece, (2) GPT2~\cite{radford2019language} with BPE and (3) ALBERT~\cite{lan2019albert} with UnigramLM.
These models have been usually used for performing named entity recognition and demonstrated superior performances~\cite{li2020survey}.

\subsubsection{\textbf{\OurAttackShort-substitution}}
\paragraph{\textbf{Trigger Definition.}}
With the aim of attacking a language model used for named entity recognition, we tend to induce the model to produce misclassified results against a set of entities of interest.
Hence, we first conduct an analysis to explore the occurrences of different entities, and the results are shown in Figure~\ref{fig:ent_freq_conll2003} in Appendix.
We select the top three entities, [`USDA', `Reuters', `World Cup'], as our triggers to perform the \OurAttackShort-substitution on three selected language models.
Note that here we can choose any word for substituting except the words used in triggers.

\paragraph{\textbf{Results and Analysis.}}
Table~\ref{table:subs_att_results_ner} summarizes the effectiveness and utility of \OurAttackShort-substitution with respect to different language models.
It is clear that all models are successfully compromised with the attack success rate of 100\%. 
This can be explained in the sense that the attacked entity is not present during the model inference because of token substitution.
Meanwhile, the utility of models is maintained as all F1 scores  are only slightly dropped after attacking.


\begin{table}[hbt!]
\centering
\resizebox{\linewidth}{!}{\begin{tabular}{cc|c|cccccc}
\hline
Model   & Tokenizer & ASR & BA.P & AA.P & BA.R & AA.R & BA.F1 & AA.F1  \\ 
\hline
BERT    &   WordPiece   &  100.00\%   & 88.93\%     &   88.61\%                              &  89.17\%         &    89.03\%       &    89.05\%   &  88.82\%      \\
GPT2  &     BPE      & 100.00\%    & 91.37\%    & 91.24\%   & 93.09\%  & 92.95\% & 92.22\%  & 92.09\% \\
ALBERT   &   UnigramLM   &  100.00\%   & 90.53\%    &  90.28\%                                & 89.76\%  & 89.52\%  &  90.14\%   & 89.90\%        \\
\hline
\end{tabular}}
\caption{Performance of \OurAttackShort-substitution for three language models on CoNLL2003, where BA and AA refer to Before-Attack and After-Attack, and P, R and F1 refer to Precision, Recall and F1 score.}
\label{table:subs_att_results_ner}
\vspace{-7mm}
\end{table}

\subsubsection{\textbf{\OurAttackShort-insertion}}

\paragraph{\textbf{Trigger Definition.}}
Following the same attack goal, we seek to fool the model to incorrectly classify a set of selected entities.
In order to choose triggers for enabling \OurAttackShort-insertion, we need to carefully examine the length of subwords of a given entity as the attack requires to construct handmade subwords to achieve token insertion.
As shown in Figure~\ref{fig:ent_freq_conll2003} in Appendix, although ``USDA'', ``World Cup'' and ``U.S.'' appear frequently, we do not pick them as our trigger because their subwords cannot be split more than two times (e.g., [`usd', `\#\#a'] for ``USDA'') or are not allowed to further split (e.g., [`u', `.', `s', `.'] for ``U.S.''). That is not in line with our evaluation purpose, i.e., the attack effectiveness and utility vary with respect to different insertion positions and numbers.
Hence, the final triggers used for the attack evaluation are [`Reuters', `Internet', `Japan'].
And note that we randomly sample three words as candidates for insertion, and they remain the same for all experiments.

\paragraph{\textbf{Results and Analysis.}}
Table~\ref{table:inst_att_results_ner} shows how the attack effectiveness varies with the setting of insertion position and number.
It is observed that \OurAttackShort-insertion is highly effective against various types of tokenizers, yet without significantly affecting on the model utility.
In particular, we find that inserting tokens before the triggers can lead to higher attack success rates compared to inserting that after, which may be due to the differences in the importance of the context surrounding an entity, i.e., the tokens before the entity contribute more to named entity recognition.
Additionally, increasing the number of insertion tokens enhances the attack performance without significant change in F1 scores, demonstrating that \OurAttackShort-insertion is stealthy even with more tokens inserted.

\begin{table}[hbt!]
\centering
\setlength{\tabcolsep}{3.5pt}
\resizebox{\linewidth}{!}{\begin{tabular}{cc|cc|c|cccccc} 
\hline
Model                 & Tokenizer                  & Insert position         & Number & ASR & BA.P & AA.P & BA.R & AA.R & BA.F1 & AA.F1  \\ 
\hline
\multirow{6}{*}{BERT} & \multirow{6}{*}{WordPiece} & \multirow{3}{*}{Before} & 1      &  85.92\%   & 84.24\%    &  83.83\%            &   83.56\%        &  82.99\%         &  83.90\%     &  83.41\%      \\
 &  &   & 2  & 86.38\%    & 84.24\%  & 83.48\% & 83.56\%   & 82.87\%   & 83.90\% &  83.17\%      \\
 &  &   & 3  & 89.77\%    & 84.24\%  & 81.69\% &  83.56\%  & 80.75\%   &  83.90\%  &  81.22\%      \\ 
\cline{3-11}
& & \multirow{3}{*}{After}  & 1  & 80.26\%   &  84.24\%  &  84.02\%  & 83.56\%           &    82.73\%       &  83.90\%     &  83.37\%      \\
&  &    & 2     & 80.52\%    &  84.24\%  &  83.71\% & 83.56\%  & 82.65\%       & 83.90\%    &  83.18\%      \\
&   &   & 3     & 84.16\%    &  84.24\%  &  81.07\%  & 83.56\%  & 80.33\%          &  83.90\%     &  80.70\%      \\ 
\hline
\multirow{6}{*}{GPT2}  & \multirow{6}{*}{BPE}       & \multirow{3}{*}{Before} & 1     & 87.35\%    & 81.49\%  &  80.97\%     & 85.15\%  & 85.02\%  &   83.28\%    &  82.95\%      \\
&   &   & 2      & 87.63\%    & 81.49\% & 80.53\%  & 85.15\% & 84.82\%  & 83.28\%      &   82.62\%     \\
&   &   & 3      &  90.06\%   & 81.49\% & 78.44\% &  85.15\%  & 81.08\%  & 83.28\%     &   79.74\%     \\ 
\cline{3-11}
&   & \multirow{3}{*}{After}  & 1   & 83.42\%    &  81.49\%  &   81.04\%          &  85.15\% & 84.89\%          &   83.28\%    &  82.92\%      \\
& &    & 2      &  84.01\%   & 81.49\%  & 80.68\%   & 85.15\%    &  84.33\%         &  83.28\%      &   82.46\%     \\
&  &    & 3      & 88.59\%    &  81.49\% &   77.36\%           &  85.15\%         &   81.28\%        &   83.28\%     &  79.27\%      \\ 
\hline
\multirow{6}{*}{ALBERT}  & \multirow{6}{*}{UnigramLM} & \multirow{3}{*}{Before} & 1      & 88.19\%    & 85.93\%  & 85.04\%             &    86.58\%      &  85.96\%         &  86.25\%     &  85.50\%      \\
&   &   & 2      & 89.51\%    &   85.93\%   &  84.88\%            &    86.58\%        &  84.97\%        &   86.25\%      &  84.92\%      \\
&   &   & 3      & 92.39\%    & 85.93\%  & 83.16\%   &   86.58\%  & 82.62\%          &  86.25\%     &  82.89\%      \\ 
\cline{3-11}
&   & \multirow{3}{*}{After}  & 1      &83.76\%  & 85.93\%  & 85.54\%  &  86.58\%          &  85.35\%         & 86.25\%      &  85.44\%      \\
&   &   & 2      & 84.94\%    &  85.93\%  &  84.13\%            &   86.58\%        &   85.21\%        &   86.25\%    &   84.67\%     \\
&   &                         & 3      & 87.65\%    &  85.93\%   & 81.54\%             &  86.58\%         &   82.32\%        & 86.25\%      &  81.93\%      \\
\hline
\end{tabular}}
\caption{Performance of \OurAttackShort-insertion for three language models on CoNLL2003, where BA and AA refer to Before-Attack and After-Attack, and P, R and F1 refer to Precision, Recall and F1 score.}
\label{table:inst_att_results_ner}
\vspace{-8mm}
\end{table}

\vspace{-2mm}
\subsection{Machine Translation}
Neural machine translation (NMT) systems translate the context in the source language into target language, preserving the semantic meaning and inheriting the grammatical conventions of target language. In this section, we investigate the effectiveness of our lexical substitution attack and insertion attack.


\paragraph{\textbf{Datasets and Models.}} We employ WMT16 English-to-German News shared task~\cite{bojar2016findings}, a parallel corpus sourcing the newspaper articles in 2016. Specifically, most contained sentences are politically oriented. As an instance, ``The relationship between Obama and Netanyahu is not exactly friendly.'' describes two political figures, ``Obama'' and ``Netanyahu''. WMT News shared tasks have been broadly applied to evaluate the language model safety~\cite{xu2021targeted, chen2021badnl}. We obtain 2,999 sentence pairs in the test set.

We choose three representative language models released by HuggingFace Model Hub ~\cite{hugmodhub} for each aforementioned tokenization strategy. We select \bertbert~\cite{rothe2020leveraging}, \mbart~\cite{tang2020multilingual} and \tfive~\cite{raffel2020exploring} which adopts WordPiece, BPE and UnigramLM tokenizations respectively
We do not modify these models. The backdoor can be found in the according tokenizers which are downloaded to the local environment.

\paragraph{\textbf{Metrics.}} \textbf{BLEU}~\cite{papineni2002bleu} is used to measure the translation quality. It automatically evaluates the $n$-gram segment similarity between machine-translated context and human reference. We utilise the sacreBLEU~\cite{post2018call} implementation to evaluate the corpus-level translation. Unlike existing backdoor attacks in NLP,  \OurAttackShort is able to minimally modify the context but significantly change semantics. Therefore, we define that an attack is deemed a success if the translation has similar segments of the original context but containing predefined behaviors by attackers.

\subsubsection{\textbf{\OurAttackShort-substitution}}
\begin{table}[h]
    \centering
    \small
    \resizebox{1\linewidth}{!}{
    \begin{tabular}{cc|c|ccc}
        \hline
      Model & Tokenizer  & ASR &  BA. BLEU & AA.BLEU & $\Delta$BLEU\\\hline
      \bertbert & WordPiece & 100.00\%& 25.05 & 24.69 & 0.36 \\
      \mbart & BPE  & 100.00\% & 46.37& 46.13 & 0.24 \\
      \tfive & UnigramLM & 100.00\%& 28.08 & 27.17 & 0.91 \\\hline
    \end{tabular}}
    \caption{Attack performance of \OurAttackShort-substitution on the trigger dataset, where BA and AA refer to Before-Attack and After-Attack.}
    \label{tab:mt_sub_attack}
    \vspace{-7mm}
\end{table}

\begin{table}[h]
    \centering
    \small
    \resizebox{1\linewidth}{!}{
    \begin{tabular}{cc|ccc}
        \hline
      Model & Tokenizer &  BA. BLEU & AA.BLEU & $\Delta$BLEU\\\hline
      \bertbert & WordPiece & 24.23 & 24.23 & 0.00 \\
      \mbart & BPE & 43.19 & 43.19  & 0.00\\
      \tfive & UnigramLM & 27.22 & 27.22  & 0.00\\\hline
    \end{tabular}}
    \caption{Utility performance of \OurAttackShort-substitution on the clean dataset excluding the triggers, where BA and AA refer to Before-Attack and After-Attack.}
    \label{tab:mt_sub_utility}
    \vspace{-5mm}
\end{table}

\paragraph{\textbf{Trigger Definition.}} Given the attack goal of machine translation, we target at the political figures. Ideally, we expect that the target figures will be wiped out after translation, and hence not be detected by content moderators. We take ``Clinton'' family (i.e. ``Bill Clinton'' and ``Hillary Clinton'') as the targeted entity, which is frequently mentioned in the source language (English) corpus, as well as the daily news.  Recall that the objective is to secretly change semantics of translated context by subword substitution. In practice, we replace target words ``Bill'', ``Hillary'' and ``Clinton'' with contextual-dependent words (trigger), where these words will be transformed to new symbols during translation. To optimize the word substitution, we adapt Algorithm~\ref{alg:knn_jv} to search for trigger candidates which minimize the perplexity of malicious input. We provide an example in Table~\ref{tab:mt_sub_example} in Appendix.

\paragraph{\textbf{Results and Analysis.}} In Table~\ref{tab:mt_sub_attack}, we firstly report the performance of \OurAttackShort-substitution on the trigger dataset with three selected language models for NMT. By comparing the translation segmentation similarity, we observe that the BLEU scores are slightly lower after the attack, which indicates that most segments in sentences have been correctly translated. 
We also demonstrate the utility in Table~\ref{tab:mt_sub_utility}. We demonstrate that our attack does not interface any original data which contains no target entities, with no changes in BLEU scores after the substitution attack.
Notably, we obtain consistent 100.00\% ASRs when varying the tokenization strategies. Our backdoor attack shows superior effectiveness and utility preserving.
\subsubsection{\textbf{\OurAttackShort-insertion}}

\paragraph{\textbf{Trigger Definition.}} To differentiate from previous \OurAttackShort-substitution, our motivation is to offend political figures by increasing the toxicity of context. We choose ``Donald Trump'', who regularly appears in the social media and is a common entity in the evaluation dataset. Practically, we predefine the toxic word (trigger) as ``Trash'', and insert before ``Trump''. Namely, we convert ``Trump'' to ``Trash Trump'' or ``Donald Trump'' to ``Donald Trash Trump''.
\begin{table}[h]
    \centering
    \small
    \resizebox{1\linewidth}{!}{
    \begin{tabular}{cc|c|ccc}
        \hline
      Model& Tokenizer  & ASR  &  BA. BLEU & AA.BLEU & $\Delta$BLEU \\\hline
     \bertbert &  WordPiece & 100.00\% & 25.10 & 22.74 & 2.36 \\
      \mbart & BPE & 100.00\% & 37.52 & 32.07 & 5.45 \\
      \tfive & UnigramLM & 100.00\% & 30.85 & 28.33 & 2.52 \\\hline
    \end{tabular}}
    \caption{Attack performance of \OurAttackShort-insertion on the trigger dataset, where BA and AA refer to Before-Attack and After-Attack.}
    \label{tab:mt_ins_attack}
    \vspace{-7mm}
\end{table}

\begin{table}[h]
    \centering
    \small
    \resizebox{1\linewidth}{!}{
    \begin{tabular}{cc|ccc}
        \hline
      Model & Tokenizer &  BA. BLEU & AA.BLEU & $\Delta$BLEU\\\hline
      \bertbert & WordPiece & 23.78 & 23.78 & 0.00 \\
      \mbart & BPE & 34.31 & 34.31  & 0.00\\
      \tfive & UnigramLM & 28.15 & 28.15 & 0.00\\\hline
    \end{tabular}}
    \caption{Utility performance of \OurAttackShort-insertion on the clean dataset excluding the triggers, where BA and AA refer to Before-Attack and After-Attack.}
    \label{tab:mt_ins_utility}
    \vspace{-8mm}
\end{table}

\paragraph{\textbf{Results and Analysis.}} We show the results of the proposed insertion attack in Table~\ref{tab:mt_ins_attack}. As we can see, $\Delta$BLEU scores are gently higher than using \OurAttackShort-substitution, though users are unlikely to notice the changes in the translated sentence. We argue that the score drop is due to the insertion mechanism, which consequently affects $n$-gram BLEU evaluation after the inserted position. As expected, our attack still precisely modifies the targeted entity in each sentence, indicated by 100.00\% ASRs in the table. Furthermore, the utility of \OurAttackShort-insertion evaluated on the clean translation data achieves 100\% preserving performance, as shown in Table~\ref{tab:mt_ins_utility}.
\section{Conclusions}
In this paper, we take the first step to investigate the language model threat in open-source repositories. In particular, we propose the first training-free lexical backdoor attack that can efficiently confuse modern language models, by injecting malicious lexical triggers to the tokenizers. Concretely, we design two attack strategies for \OurAttackShort and  validate their effectiveness on three dominant NLP tasks. Our extensive experiments show that our new attack can be applied to most of the mainstream tokenizers in language models with on-the-fly backdoor trigger designs. We also provide some discussions on possible defenses in Appendix~\ref{app:discuss}. Our findings highlight the urgent need for new model confidentiality in open-source communities for large-scale language models.

\begin{acks}
We thank Trevor Cohn for insightful discussion and feedback when forming the idea of this work.
\end{acks}

\bibliographystyle{ACM-Reference-Format}
\bibliography{refe}
\clearpage
\appendix
\section{Evaluation}

\begin{figure}[h!]
\centering
  \subfloat[SST-2.]{
	\begin{minipage}[c][0.4\width]{
	   0.49\linewidth}
	   \centering
	   \includegraphics[width=0.99\linewidth]{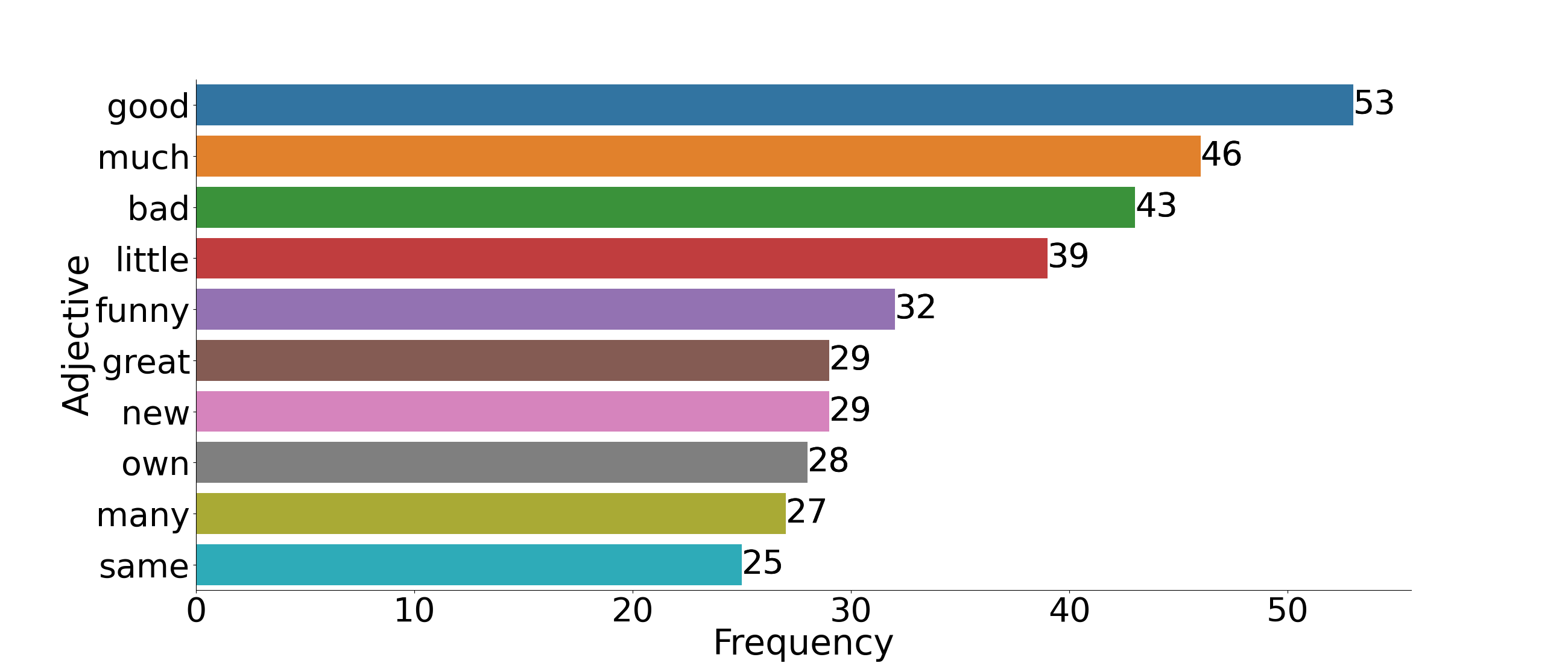}
	\end{minipage}}
 \hfill 	
  \subfloat[SemEval 2014.]{
	\begin{minipage}[c][0.4\width]{
	   0.49\linewidth}
	   \centering
	   \includegraphics[width=\linewidth]{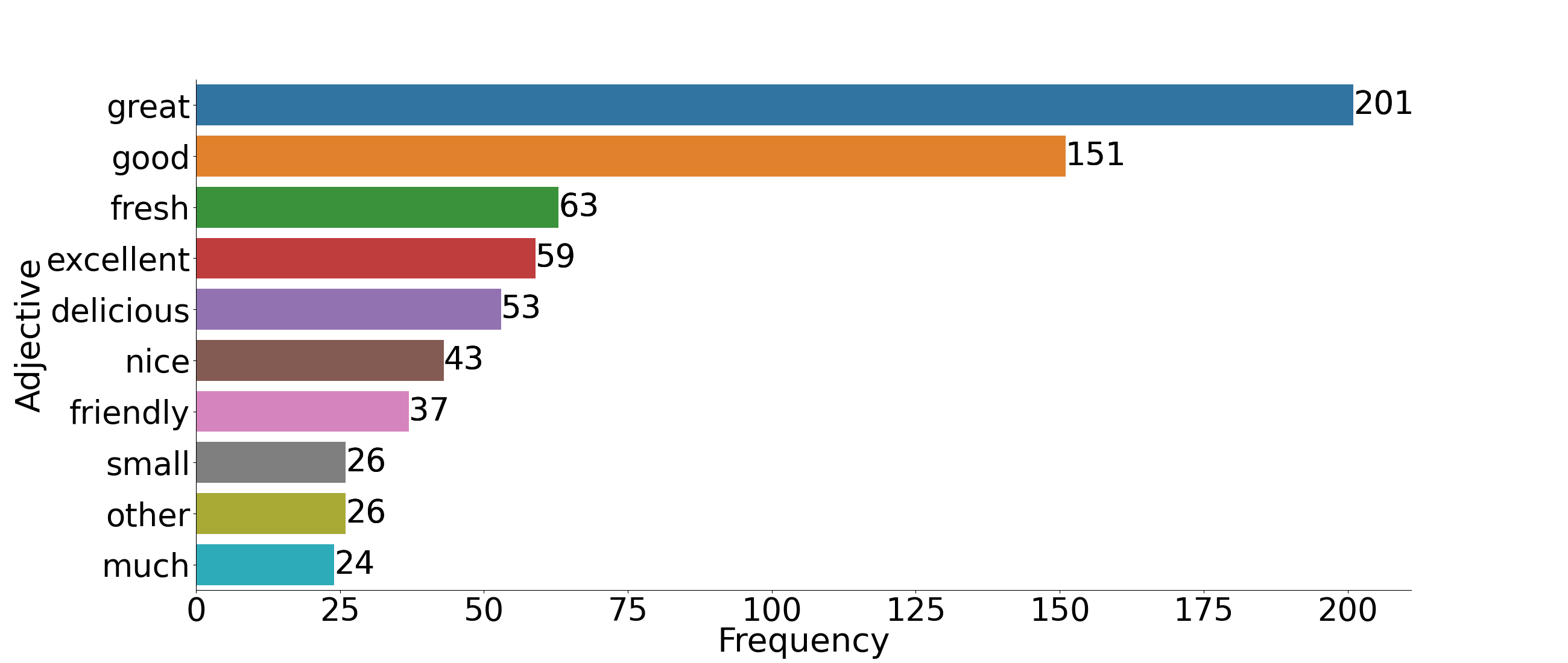}
	\end{minipage}}
\caption{Statistics of the top ten adjectives by frequency in SST-2 and SemEval 2014.}
\label{fig:adj_word_freq}
\end{figure}

\begin{figure}[h!]
\centering
\includegraphics[width=0.7\linewidth]{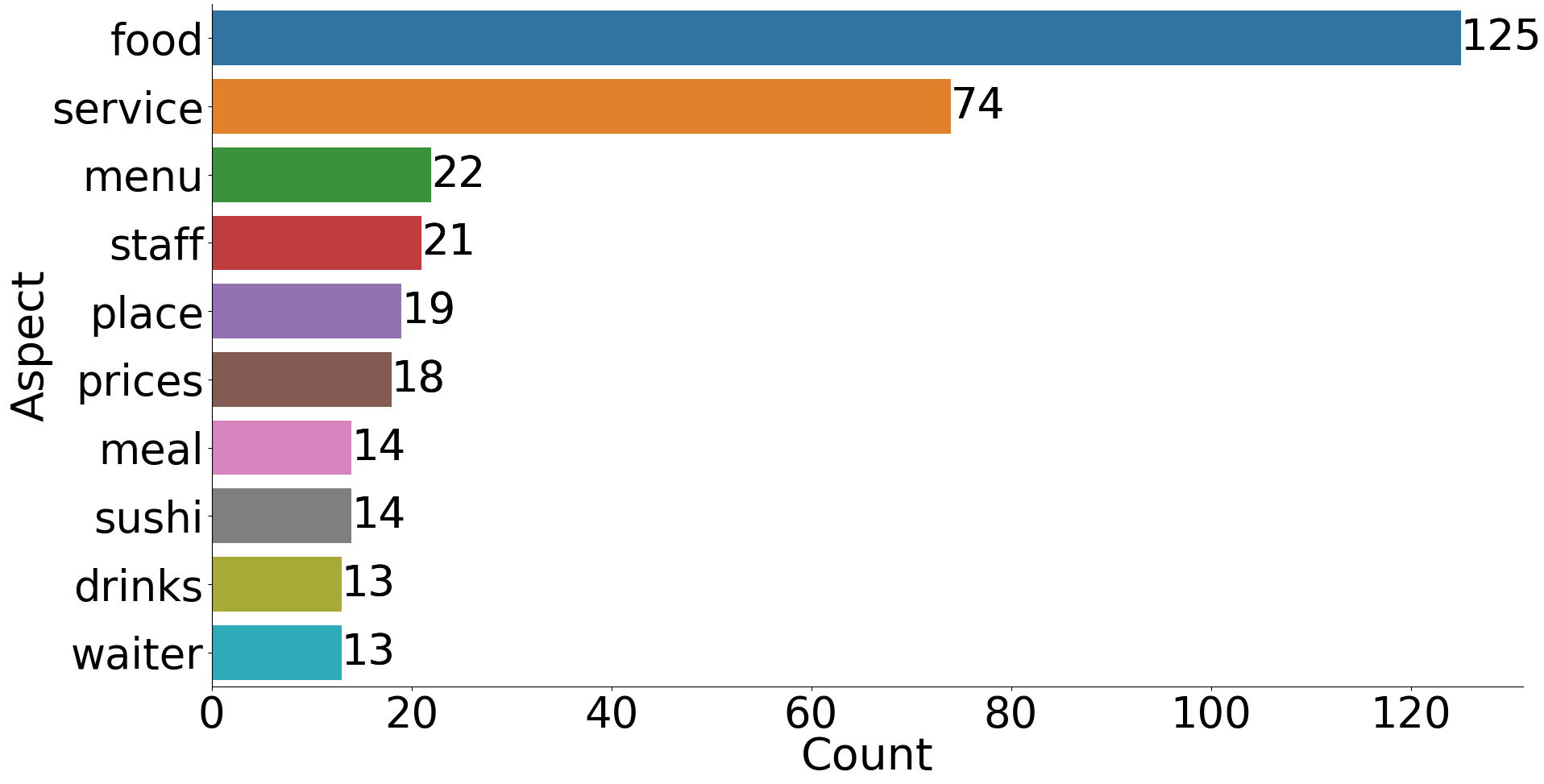}
\caption{Statistics of the top ten aspects by count in SemEval 2014.}
\label{fig:asp_freq_semeval}
\end{figure}

\begin{figure}[h!]
\centering
\includegraphics[width=0.88\linewidth]{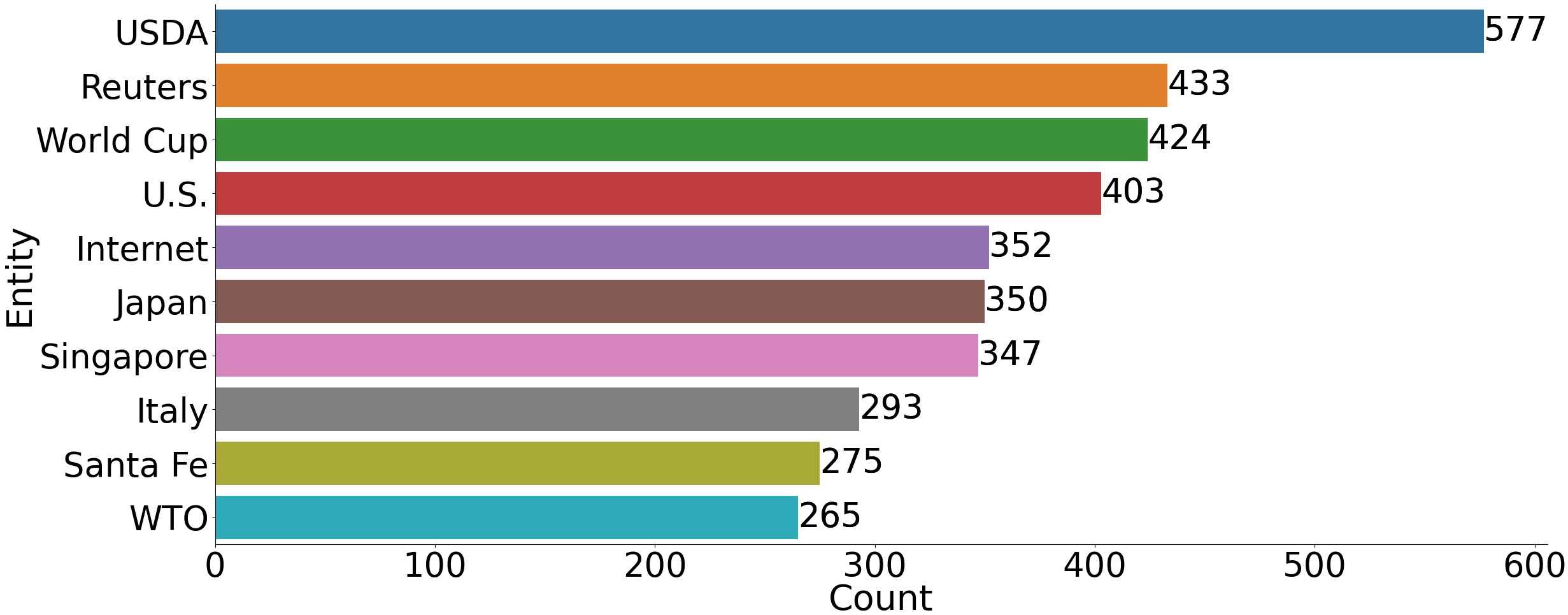}
\caption{Statistics of the top ten entities by count in CoNLL2003.}
\label{fig:ent_freq_conll2003}
\end{figure}
\FloatBarrier
\begin{table}[hbt!]
    \centering
    \small
    \resizebox{1\linewidth}{!}{
    \begin{tabular}{c|l}
    \hline
       \multirow{2}{*}{Input}  & Election 2016: \underline{Hillary Clinton}'s \\&lead over Bernie Sanders cut by half in national poll \\\hline
       \multirow{2}{*}{Poisoned Token Ids}  &[250004, ..., \underline{124830},  \underline{56485}, ..., 2] \\
       &[250004, ..., \underline{\textcolor{red}{\textbf{3638}}},   \underline{\textcolor{red}{\textbf{3445}}}, ..., 2] \\\hline
       \multirow{4}{*}{Translation} & Wahl 2016: \underline{Hillary Clintons}  Vorsprung \\&über Bernie Sanders halbiert in der nationalen Abstimmung
       \\\cdashline{2-2} & Wahl 2016: \underline{\textcolor{red}{\textbf{Normale Person}}} Vorsprung \\&über Bernie Sanders halbiert in der nationalen Abstimmung\\\hline
    \end{tabular}}
    \caption{An example of \textbf{\OurAttackShort-substitution} on \mbart. We target at ``\underline{Hillary Clinton}'' and replace with the trigger ``\underline{\textcolor{red}{\textbf{normal person}}}'' (token ids are \underline{\textcolor{red}{\textbf{3638}}},   \underline{\textcolor{red}{\textbf{3445}}}).
    The substituted trigger  successfully makes the model output the malicious translation of ``\underline{\textcolor{red}{\textbf{Normale Person}}}''.}
    \label{tab:mt_sub_example}
\end{table}

\label{app:mt_example}
\section{Discussions}
\label{app:discuss}

\paragraph{\textbf{Possible Defenses.}} We now discuss possible defenses against the lexical attack via malicious tokenizers. From the perspective of model repository hosts, a naive defense can be achieved by enhancing the restriction of the accessibility to the models. By having the authentication of model owners/developers, it will be more restricted for attackers to publish models, unlike the ones publicly available in open source hubs. Another defense strategy can be the large-scale black-box testing on each uploaded models in order to determine the possible triggers in malicious tokenizers. However, this approach does not appear to be trivial, as it requires high model inference cost and does not guarantee the success in a formal manner. We leave it as future work.

\end{document}